# Exploring the Surface Segregation of Rh Dopant in PtNi Nanoparticles through Atom Probe Tomography Analysis

Se-Ho Kim [a,b,*], Hosun Jun [c], Kyuseon Jang [c], Pyuck-Pa Choi [c], Baptiste Gault [a,d], Chanwon Jung [c,*]

[a] Max-Planck-Institut für Eisenforschung, Max-Planck-Straße 1, 40237 Düsseldorf, Germany

[b] Department of Materials Science and Engineering, Korea University, Seoul 02841, Republic of Korea

[c] Department of Materials Science and Engineering, Korea Advanced Institute of Science and Technology (KAIST), 291 Daehak-ro, Yuseong-gu, Daejeon 34141, Republic of Korea

[d] Department of Materials, Imperial College London, SW7 2AZ London, UK

* Corresponding Authors: sehonetkr@korea.ac.kr, c.jung@mpie.de

## Abstract

Proton exchange membrane fuel cells hold promise as energy conversion devices for hydrogen-based power generation and storage. However, the slow kinetics of the oxygen reduction at the cathode imposes the need for highly active catalysts, typically Pt or Pt-based, with a large available area. The scarcity of Pt increases deployment and operational cost, driving the development of novel highly-active material systems. As an alternative, Rh-doped PtNi nanoparticle has been suggested as promising oxygen reduction catalyst, but the 3D distributions of constituent elements in the nanoparticles have remained unclear, making it difficult to guide property optimization. Here, a combination of advanced microscopy and microanalysis techniques is used to study the Rh distribution in the PtNi nanoparticles, and Rh surface segregation is revealed, even with an overall Rh content below 2 at. %. Our findings suggest that doping and surface chemistry must be carefully investigated to establish a clear link with catalytic activity can truly be established.

Research on hydrogen-generation is intensifying to facilitate the transition to net zero carbon emissions. Proton exchange membrane fuel cells (PEMFCs) that convert chemical energy of hydrogen into electrical energy through hydrogen oxidation at the anode and oxygen reduction at the cathode are considered one of the future pillar of the hydrogen economy[1,2]. PEMFCs have considerable potential in transportation as a substitute for combustion engines, and have already been adopted[3]. However, due to the slow reaction rate of oxygen reduction at the cathode, implementation of a large amount of Pt catalyst is unavoidable. The scarcity and cost of Pt drives a strong effort to design less expensive catalysts from Earth abundant elements, while maintaining high activities[4–7].

A common design approach to enhance catalytic activity and reduce the use of costly Pt, which in turn can modify the oxygen binding energy of the catalysts, is by alloying with transition metals such as Fe[8,9], Co[7,10,11], Ni[12–15], and Mn[16,17]. For instance, Pt-Ni alloy catalysts exhibit a substantial improvement in mass activity up to a factor of 10 compared to commercial Pt/C[4]. However, it has been found that alloy systems of Pt-M nanoparticles have poor durability, because the transition metal dissolves in the acidic solution under the operation condition of a PEMFCs[8–10]. To address this drawback, adding a small amount of a third element, i.e. a strategy referred to as doping, has been suggested, in order to improve both durability and oxygen reduction activity. Since the successful demonstration of the Mo-doped Pt-Ni catalyst in 2015, which have shown 80 times higher catalytic activity than Pt-Ni and ~35% improved durability[14], research on ternary element catalysts based on Pt-Ni has been actively pursued, and increasingly more complex systems are being explored, with e.g. octonary alloys[18,19], to further advance catalysts.

For characterization of multi-component nanocatalysts, scanning transmission electron microscopy-energy dispersive X-ray spectroscopy (STEM-EDS) is commonly used for elemental mapping of nanocatalysts, and is increasingly complemented by atom probe tomography (APT)[20–22]. APT has the ability to measure elements distribution in three-dimensions with high chemical sensitivity and sub-nm spatial resolution, and, as such can address issues associated to the two-dimensional projected imaged obtained by STEM-EDS that makes analysis of in-depth compositional distribution in 3D challenging[23]. APT also exhibits an equal sensitivity to both light and heavy elements, whereas STEM-EDS is relatively less sensitive to light elements[24].

Here, we analyzed Rh-doped PtNi nanoparticles that have been reported to exhibit excellent oxygen reduction activity and superior stability[25–27]. We conducted a comprehensive comparison of APT with the conventional STEM-EDS on these nanoparticles to highlight the complementarities. Our analysis delves into the surface chemistry of these nanoparticles, offering potential insights into the impact of trace elements on the catalytic properties of nanoparticles.

The Rh-doped PtNi nanoparticles were first synthesized following the protocol described in Ref. [25], at room temperature and using high-quality analytical grade chemicals from Sigma Aldrich. TEM images in Figure 1a reveal as-synthesized, uniformly-shaped octahedral nanoparticles with a size of 8.2 ±0.8 nm. At higher magnification, Figure 1b, sets of {111} planes are imaged with an interplanar distance of 2.22 ±0.02 Å[25]. STEM-EDS shows 67.2 ±8.0 at.% Pt and 32.8 ±8.0 at.% Ni. In earlier investigations on heavily Rh-doped (11 at.%) $Pt_3Ni$ nanoparticles[25], an accumulation of Rh atoms at the surfaces of the ternary nanoparticles was observed. However, while Pt and Ni peaks were clearly observed in the spectrum in Figure 1c, the Rh-L$\alpha$ peak at 2.696 eV was barely noticeable.

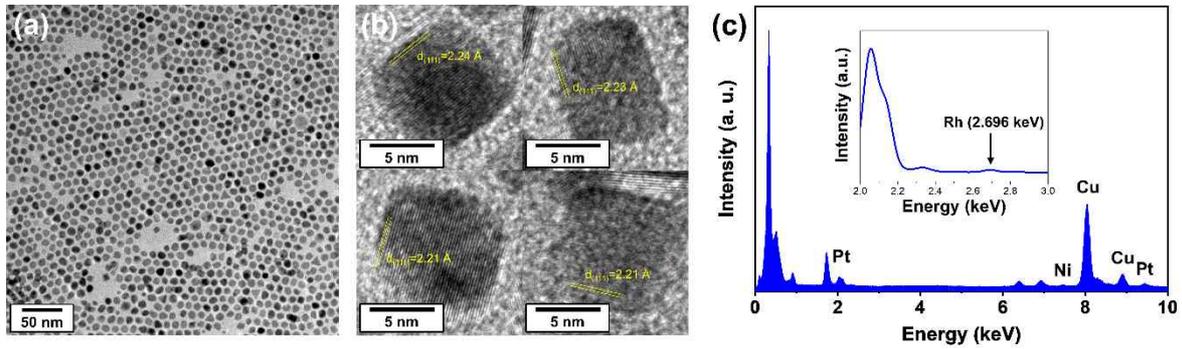

**Figure 1.** TEM image of as-synthesized Rh-doped PtNi nanoparticles with (a) low magnification and (b) high magnification. (c) EDS spectra of Rh-doped PtNi nanoparticles (inset: enlarged region with 2-3 keV). Note that Si peak comes from the detector while signals of Fe and Co originate from contamination on the detector.

For APT measurement on nanoparticles, we introduced co-electrodeposition, an enabling versatile approach to embed freestanding nanoparticles in a metallic matrix and facilitate preparation of specimens[28] by focused-ion beam milling. Here, the as-synthesized nanoparticles were deposited first by electrophoresis on a pure Co substrate. The residual solution is removed. Scanning electron microscopy (SEM) images of the as-synthesized Rh-doped PtNi nanoparticles deposited on the Co substrate in Figure 2a–b revealed that the nanoparticles remained intact and were not subjected to deformation nor corrosion during the electrophoretic process.

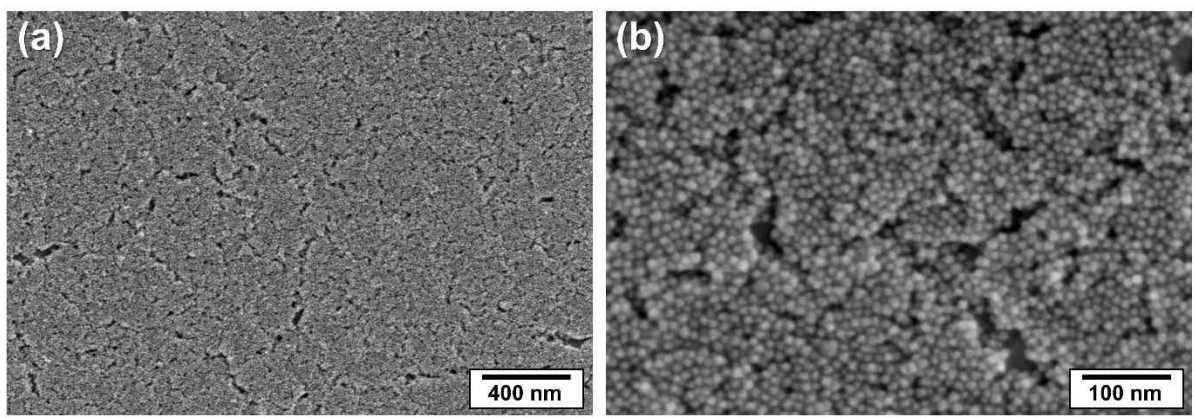

**Figure 12.** (a,b) SEM images of anode substrate with electrophoresed Rh-doped PtNi nanoparticles.

After electrophoretic deposition of Rh-doped PtNi nanoparticles, 100 cycles of cyclic voltammetry (CV) between 0 and 1.0 V (vs. RHE) in 0.5 M NaOH with scan rate of 0.5 V s$^{-1}$ was performed on the electrode to remove the ligands used for nanoparticle synthesis[29]. The solution was then substituted with a solution containing Co ions for electroplating to encapsulate the nanoparticles. Pulsed electrodeposition was then carried out to fill the voids between the particles with Co, using a current of 5 mA for 120 s[30]. Figure 3a shows a cross-sectional STEM image of the encapsulated nanoparticles in the Co film, with a high concentration of nanoparticles at the surface of the Co-substrate. Based on the elemental mappings in Figure 3b, it appears that no element has undergone galvanic dissolution into the Co matrix during electrodeposition. However, the Rh signal in EDS remained too weak for precise quantitative analysis. As an indication, the regions with a high particle density comprised 65.5 ±6.8 at. % Pt, 32.2 ±0.4 at. % Ni, and 2.4 ±0.1 at. % Rh.

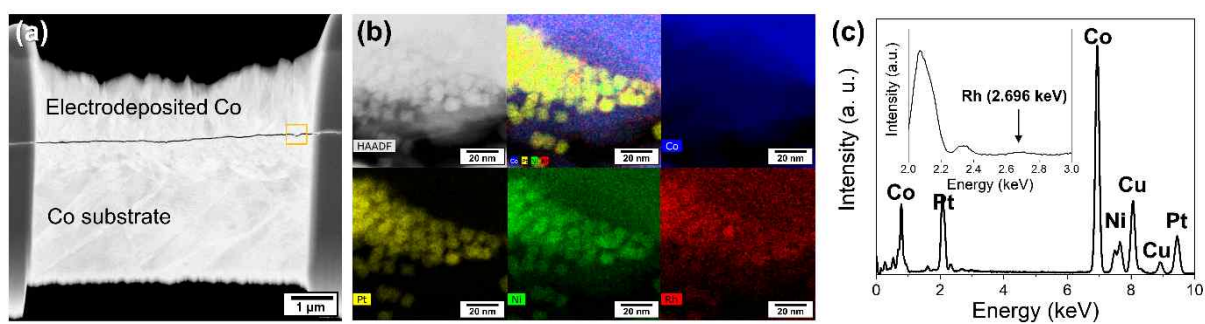

**Figure 2<s>3</s>.** (a) Cross-sectional STEM image after pulsed electrodeposition. (b) EDS mapping and (c) the corresponding spectrum of dense particle zone.

The APT measurement was performed in UV-laser mode in 4000 HR LEAP at a base temperature of 45.5 K, a pulsing frequency of 125 kHz, and a detection rate of 0.5 %. AP Suite provided by CAMECA was used to reconstruct a dataset of 7.3 × 10$^6$ atoms of Pt, Ni, and Rh obtained from a set of agglomerated nanoparticles, i.e. excluding the Co-matrix atoms. Considering atom-detection efficiency of 37 % of 4000 HR atom probe model, 2 × 10$^5$ atoms of Pt, Ni, and Rh elements were collected. Thus, each 8-nm-length octahedral face-centered cubic nanoparticle is calculated to consist of approx. 30,000 atoms indicating that there are ~7 nanoparticles in this dataset. The 3D atom map of the Rh-doped PtNi nanoparticles embedded in Co is displayed in Figure 4a. The iso-compositional surface at 33 at.% Pt (yellow) highlights the interface between the nanoparticle and the matrix, and it is the corner-like shape which suggests an edge of an octahedral nanoparticle based on the result of TEM imaging. Co atoms (shown in blue) is not expected within the nanoparticles, however the overlap in the mass spectrum between the tail of the main Ni peak and Co makes it difficult to precisely assess. The overall composition for Pt, Ni, and Rh is 64.8 ±0.2, 33.5 ±0.3, and 1.7 ±0.4 at.%, respectively. Pt and Ni atoms appear randomly distributed in the nanoparticle, whereas Rh atoms, which are displayed as red spheres, show a tendency for agglomeration evidenced in the 3-nm slice through the tomogram in Figure 4b. The elemental distribution in the nanoparticles is further quantified by using one-dimensional compositional profiles positioned normally to the surface of individual nanoparticles, Figure 4c and 4d, indicating up to 4–6 at% of Rh segregation on the nanoparticles' surfaces.

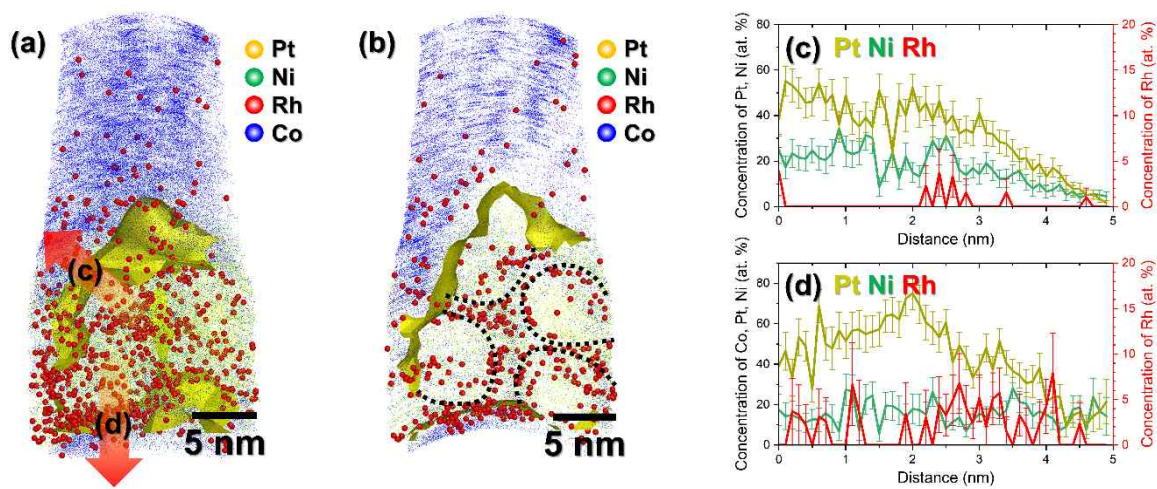

**Figure 34.** (a) 3D atom map for all the elements including Pt 20 at.% iso-surfaces (b) 3-nm thin sliced Rh-doped PtNi nanoparticles embedded in Co. Red dots represent the reconstructed Rh atoms and dotted black lines represent each nanoparticle surface. (c) 3D atom maps of constituent elements (d) 1D compositional profile (ϕ5 × 5 nm³) across the interface between Rh-doped PtNi NPs and Co matrix.

Challenges in quantifying trace amount of Rh segregation in ternary nanoparticles have precluded their study, leading to an erroneous assumption that Rh was homogenously distributed. For instance, Strasser et al. studied $Pt_3Ni$ nanoparticle doped with 5 at.% Rh revealing surface segregation of Rh while in the case of 4 at.% Rh-doped nanoparticles, it was difficult to differentiate the segregation of Rh using STEM-EDS. This finding suggests that it is challenging to assess the degree of surface of bulk segregation below a concentration threshold, here in the range of 4 at%, i.e. higher than in the particles we synthesized in which we measured 1.7 ±0.4 at.% Rh by APT.

Recent experiments and computational calculations on bulk Pt-Rh suggest that there is no miscibility gap [31], contradicting the phase diagram reported by Raub [32]. In nanomaterials, distinct surface segregation of Rh has often been observed in the Pt-Rh system due to external driving forces by surface adsorbents, including $O_2$ and $H_2$ from the atmosphere[33,34]. Although the synthesis of the nanoparticle in this study was carried out under a $N_2$ environment, exposure to air during washing and storage was unavoidable and this could have triggered surface segregation of Rh. On the contrary to Pt-Rh, the Ni-Rh system has a miscibility gap close to the Rh-rich side of the phase diagram[35]. Surface energy, strain energy, and bond energy induce element segregation of Rh, and Su et al. have demonstrated why the trace amount of Rh is enough to increase the stability for Ni-Rh nanoparticles[36]. Studying how the distribution of minor elements, like Rh, changes under conditions of oxygen reduction reaction (ORR) could be a promising area of research for future investigations.

In summary, we synthesized Rh-doped PtNi nanoparticles, and revealed that, contrary to earlier suggestion that Rh atoms are homogeneously distributed inside the nanoparticles, Rh segregates to the surface. Our finding evidence that the chemistry of the nanoparticle is more complex than previously reported, and investigation the relationship between composition and catalytic activity (e.g. for the ORR) must be revisited to account for this newly revealed surface chemistry, which will form the basis for future studies.

## Acknowledgment

This work was supported by the National Research Foundation of Korea (NRF) (grant number 2020R1A6A3A13073143). S.-H.K. and B.G. acknowledge financial support from the German Research Foundation (DFG) through DIP Project No. 450800666.